\documentclass[11pt]{article}
\usepackage{graphicx}
\usepackage[authoryear,longnamesfirst]{natbib}

\begin{document}
\baselineskip=18pt

\title{Photoacoustic detection of phase transitions at low temperatures 
in CsPbCl$_3$ crystals}

\author{Masanobu Iwanaga\footnote{E-mail: iwanaga@phys.tohoku.ac.jp} \\
Department of Physics, Graduate School of Science, \\
Tohoku University, Sendai 980-8578, Japan}


\maketitle

\begin{abstract}
\baselineskip=18pt
Phase transitions in CsPbCl$_3$ crystals have been explored 
by photoacoustic (PA) method in a wide temperature range of 150--300 K. 
The PA signals, induced by ns-pulsed laser excitation of electronic states 
and detected with a piezoelectric transducer, are thermoelastic ultrasonic 
waves lasting for hundreds of microseconds. 
The wavelet analysis of the PA wave has clarified that the thermoelastic 
wave includes several MHz components. 
The PA signals have been made use of 
in order to examine phase transitions at low temperatures in the crystal. 
Changes of PA signals indicate that two phase transitions 
take place at low temperatures around 200 K. The types of phase 
transitions are discussed taking into recent experimental results. 
\end{abstract}
{\it Keywords:} Photoacoustic spectroscopy; Ultrasonic wave; 
Phase transitions; CsPbCl$_3$

\newpage
\section{INTRODUCTION\label{intro}}
CsPbCl$_3$ crystal was investigated as a typical material undergoing 
structural phase transitions above room temperature (RT) \citep{Fujii}; 
the crystal is perovskite in the temperature range above 320 K, 
transforms to tetragonal at 320 K, and becomes orthorhombic or monoclinic 
at 310 K. The phase transitions are considered to be ferroelastic, related 
to zone-boundary phonons of M$_3$ and R$_{25}$ \citep{Fujii,Hirotsu}.  

On the other hand phase transition below RT was 
explored in the measurement of doped CsPbCl$_3$ with the 
electron-paramagnetic-resonance (EPR) method, so that the results implied 
that phase transition takes place at about 180 K \citep{Cape,Cohen}. 
However the temperature of phase transition estimated from the EPR 
experiments is 185 \citep{Cape} and 176 K \citep{Cohen}, 
and slightly different with each other. 
The discrepancy could come from the doping effect on the phase transition, 
and therefore it seems more appropriate to examine the non-doped crystal 
for determining the phase-transition point. 
The crystal structure below RT has been studied recently with 
the x-ray diffraction \citep{Sakuma}; the study concluded that the crystal is 
orthorhombic below RT and implied that structural 
transition does not take place below RT. Thus phase transition 
below RT is not conclusive from the structural examination. 
The result implies that if phase transition 
does take place below RT, it can be second-order transition 
that is not associated with any crystallographic transition. 

Another recent study on phase transition at low temperatures has 
been performed by focusing on 
optical absorption tail and photoluminescence (PL) of CsPbCl$_3$. 
Since the crystal is a semiconductor with direct 
bandgap \citep{Heidrich}, the electronic excited states are generated 
under photoexcitation. The absorption tail and PL properties 
suddenly change around 170 K \citep{Hayashi}; the results suggest 
phase transition below RT. 

To explore the phase transition, photoacoustic spectroscopy (PAS) 
is a suitable method because PAS enables us to observe self-induced 
thermoelastic ultrasonic wave under photoexcitation. In fact phase 
transitions have been often studied by PAS and detected 
successfully \citep{Jackson,Etxe}; even 
second-order transition were observed. 
In the present study, it is aimed 
to clarify the properties of observed PA signals and furthermore to test 
the existence of phase transition below RT. 

Experimental details are described in Sec.\ \ref{exp}. 
Experimental results are presented and the wavelet analysis is applied 
to the PA wave in Sec.\ \ref{results}. 
The wavelet analysis has the advantage of resolving the 
superimposed waves into the several waves corresponding to each frequency 
component; for example, the electric noise 
mixed in the experimental data is apparently separated from PA signals. 
After clarifying PA-signal properties, the existence of phase transitions 
below RT is discussed in Sec.\ \ref{discussion}. 

\section{EXPERIMENTAL DETAILS\label{exp}}
CsPbCl$_3$ crystals were grown with the Bridgman technique. To purify 
99.999\% PbCl$_2$ powder, the single crystal was previously grown, 
and 99.99\% CsCl powder and the PbCl$_2$ single crystal were mixed at 
equimolar amount and sealed in a quartz ampoule at about 10$^{-5}$ Torr. 
The ampoule was set in a Bridgman furnace and was moved from the point 
of 970 K to that of 720 K at about 20 K/day. 
Though the melting temperature of CsPbCl$_3$ is about 880 K, 
the ampoule was kept below the melting point for several days 
to anneal the specimen. Finally it was cooled from 720 K to RT for a day. 
The as-grown ingot of CsPbCl$_3$ was aggregate of single crystals. 

\begin{figure}
\begin{center}
\includegraphics[scale=.7]{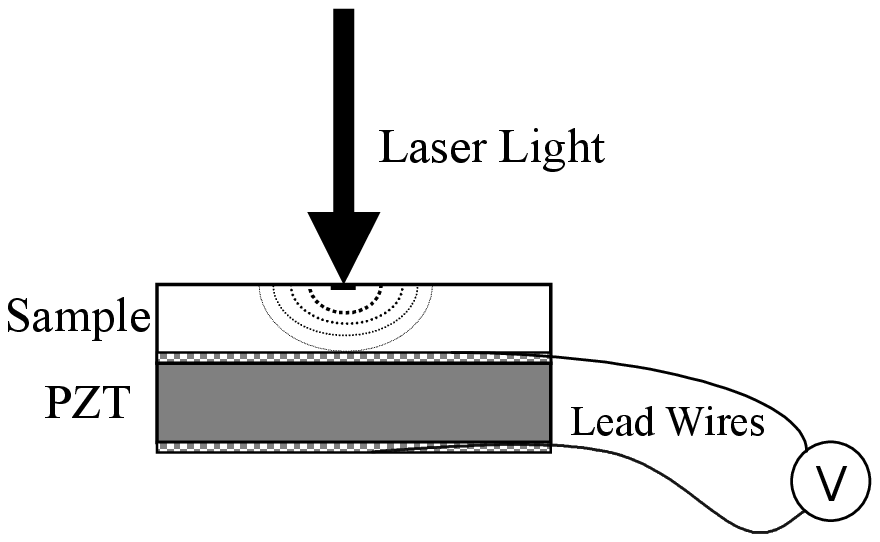}%
\end{center}
\caption{\baselineskip=20pt
Schematic configuration of a CsPbCl$_3$ and a PZT. 
The sample was photoexcited 
on one cleaved plane and PA signals were detected on the other by the PZT 
attached with conductive organic paste. Dotted lines in the sample stand for 
the PA waves schematically. The piezoelectric voltage was 
measured along the thickness direction. This figure is drawn for the PZT 
of 5$\times$5$\times$1 mm$^3$. More details are described in Sec.\ \ref{exp}.}
\label{fig1}
\end{figure}

The crystal was cleaved in $ab$ plane perpendicular to $c$ axis; 
the crystals can be cleaved only in $ab$ plane. Typical dimension of the 
cleaved specimen was 5$\times$5$\times$1 mm$^3$ along $a$, $b$ and $c$ axes 
respectively. Both cleaved $ab$ planes were used for PA measurement; 
one was the plane injected by incident photons and the other was firmly 
attached to a piezoelectric transducer (PZT) with conductive organic 
paste. Experimental configuration 
is schematically drawn in Fig.\ \ref{fig1}. 

Two type of PZT were used; the PZT of 3.99-MHz resonant 
frequency was a plate of 5$\times$5$\times$1 mm$^3$ and 
that of 1.97-MHz resonant frequency was a 10-mm diameter and 1-mm 
thickness disc. One of the cleaved planes was 
attached to a PZT as shown in Fig.\ \ref{fig1}. 
Curie temperatures of the PZTs were 603 K and therefore 
the PZTs do not undergo any phase transition in the temperature range of 
150--300 K at which the present measurement was performed. 
The crystal and PZT were set in a He-flow cryostat. 
Acoustic-wave ringing effect in a PZT was checked; the effect 
is compared with the PA wave propagated from CsPbCl$_3$ in the next section. 

The PA signals were induced by the third harmonic generation of Nd:YAG 
laser; the laser pulses were 3.49 eV and 5-ns width, and 
the reputation was 10 Hz. Since CsPbCl$_3$ crystals have the fundamental 
absorption edge at 3.0 eV \citep{Heidrich}, 
the incident light is absorbed due to the electronic interband 
transition, except for the surface reflection. 
The absorption takes place in the thin surface layer because of the 
absorption length less than 1 $\mu$m and the sample thickness of 1 mm. 
In this configuration 
the PA signal generated in the surface layer propagates 
through the crystal and is detected by the PZT on the back surface. 
Thus the detected PA wave is bulk wave 
propagated through the crystal; the measured PA wave does not only propagate 
along $c$ axis but also includes the off-axis wave 
which travels along $a$- or $b$-axis and 
the multiple reflection in the crystal. 
The intensity of incident laser light was about 200 $\mu$J/pulse, and 
the incident light is loosely focused at the spot size of 1-mm diameter on 
the sample surface; the sample was not damaged 
by the laser-light irradiation. 
The PA wave was measured 
by an oscilloscope without any preamplifier. Different time-range data 
were measured by varying the coupling resistance to the oscilloscope. 

The amplitude and phase of the PA signals were picked up by a two-phase 
lockin amplifier at the cycle of incident pulse. 
The reference input to the lockin detector was taken from the 
transistor-transistor-logic pulses synchronized with the laser pulses. 
In the setup the lockin detector pick up 
the signals of the same frequency with the reference; 
the amplitude is proportional to the peak intensity because 
the lockin detector automatically tunes the phase of sine curve 
to the peak position of PA signal. Measuring the PA signals by the lockin, 
the coupling resistance is required high; 
the PA wave is therefore broadened to ms-order, 
so that the lockin technique is applicable to the present 
measurement. 

At the end the relation of the pulsed PA measurement with the conventional 
PA measurement is made mention of; pulsed PA signals are generally
stronger than the conventional PA signals induced by chopper-modulated 
laser light because the laser pulses have higher photon density \citep{Patel}. 
In the conventional PA measurement using cw laser light of 3.81 eV and a 
chopper, the amplitude and phase of PA signals of CsPbCl$_3$ show 
a profile similar to those of pulsed PA signals (see Fig.\ \ref{fig4}). 
The PA signals were just weaker than the pulsed PA signals. 
Consequently there is not any special advantage to employ 
the conventional PA technique here.

\section{RESULTS AND ANALYSIS\label{results}}

\begin{figure}
\begin{center}
\includegraphics[scale=.65]{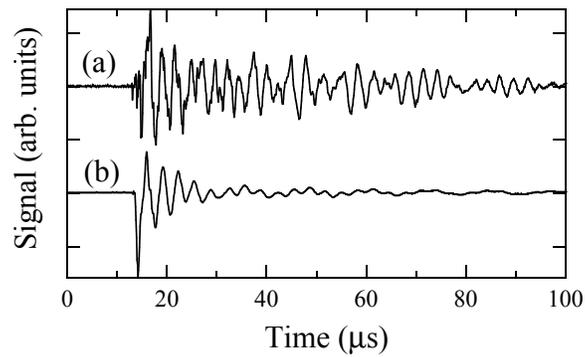}%
\end{center}
\caption{\baselineskip=20pt
(a) Detected PA wave induced by photoexcitation of CsPbCl$_3$ 
at 296 K with 3.49-eV laser pulses of 5-ns width. Incident laser pulses 
reach onto the surface at 12.5 $\mu$s. 
The PZT used in the measurement has the resonance at 3.99 MHz. 
(b) Ringing at 296 K in a PZT of 3.99-MHz resonance; it was induced by 
direct photoinjection onto the PZT. The incident light was 2.33-eV 
laser pulses of 5-ns width. 3.49-eV laser pulses also induce the same shape 
of ringing. The ringing effect on the detected PA wave is described in the 
text. Both spectra are presented to a similar scale.}
\label{fig2}
\end{figure}

\begin{figure}
\begin{center}
\includegraphics[scale=.6]{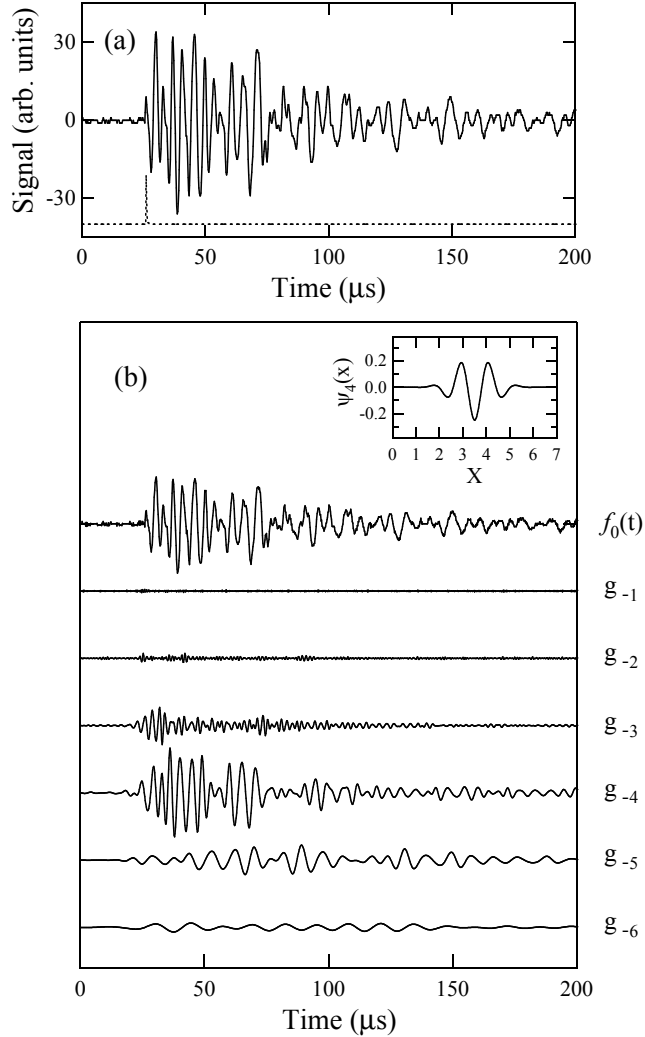}%
\end{center}
\caption{\baselineskip=18pt
(a) PA wave in CsPbCl$_3$ at 297 K 
(solid line); it is induced by 5-ns-width laser pulses of 
3.49 eV. Dashed line: the position at which the laser pulses reach onto 
the sample surface. The dashed line is vertically offset. 
(b) Wavelet analysis with the fourth B-spline wavelet $\psi_4$; 
the inset shows the $\psi_4$ which is center symmetrical. 
$f_0(t)$ denotes a good 
approximation for the PA wave in Fig.\ \ref{fig3}(a). 
$g_j$ ($j=-1, \cdots, -6$) resolve $f_0$; they 
are vertically offset and displayed to the same scale with $f_0$.}
\label{fig3}
\end{figure}

Figure \ref{fig2} shows detected PA wave at 296 K (curve (a)) in a $\mu$s 
range. The wave was induced by 3.49-eV pulsed laser light of 5-ns 
width at $t = 12.5$ $\mu$s. The PA wave continues 
to oscillate prominently for more than 80 $\mu$s. 
For comparison with the PA wave, PZT 
ringing (curve (b)) was also displayed to a similar scale; 
it was measured under direct photoirradiation onto the PZT 
attached to nothing but lead wires. 
The ringing in the PZT is directly stimulated due to the laser 
pulse, independent of photon energy of laser pulses. 
The envelope of rapid component decays exponentially at about 10 
$\mu$s and the slow component weakly survives up to 100 $\mu$s. 
As seen in Fig. \ref{fig2}, the decay of PA wave is different 
from the PZT ringing. That is, 
the ringing effect on the detected PA wave is not dominant. 
Probably the oscillatory structure of PA wave mainly comes from 
long lasting multiple reflection of the thermoelastic wave in the crystal. 
Besides the PZT ringing such as curve (b) in Fig.\ \ref{fig2} becomes 
comparable to the PA waves  
only under direct and 1.5-times more intense excitation. 
Thus, though the influence of ringing is in principal included in the 
measured PA wave, the contribution is quite small. It is also shown in the 
analysis of Fig.\ \ref{fig3}. 

Figure \ref{fig3}(a) presents the PA wave at 297 K, 
which was detected by a PZT of resonant frequency 
of 1.97 MHz. Solid line in Fig.\ \ref{fig3}(a) represents the 
measured PA wave, and dashed line shows the incident laser pulse of 3.49 eV. 

To analyse this real time oscillation, 
we choose to use the wavelet method. In the present analysis, the fourth 
B-spline wavelet $\psi_4$ in the inset of Fig.\ \ref{fig3}(b) 
is selected from many wavelets; the $\psi_4$ is a finite wave packet 
composed of third-order polynomials and corresponds mainly to a narrow 
frequency band via Fourier transformation \citep{Sakakibara}. 
First the measured PA wave in Fig.\ \ref{fig3}(a) is approximated by the 
fourth B spline. A good approximation is displayed as $f_0(t)$ in 
Fig.\ \ref{fig3}(b). The PA wave $f_0(t)$ is decomposed as follows: 
$f_0(t)=\Sigma_j g_j(t)$ 
where $g_j(t)=\Sigma_k d_k^{(j)} \psi_4(2^j t - k)$, and $d_k^{(j)}$ are 
coefficients \citep{Sakakibara}. 
The $f_0$ is well resolved by $g_j$ ($j=-1, \cdots, -6$). 
Figure \ref{fig3}(b) displays the result of wavelet decomposition. 
$g_{-1}$ and $g_{-2}$ make small contribution; the $g_{-2}$ represents 
electric noise at 25 $\mu$s which appears without PA signal and 
stems from the trigger signal for measurement. 
$g_{-4}$ is apparently predominant, and $g_{-3}$ and $g_{-5}$ 
contribute to some extent. Though $g_{-6}$ also contributes slightly, 
it almost corresponds to the PZT ringing observed in this time range. 
$g_{-3}$, $g_{-4}$ and $g_{-5}$ are consequently regarded as the contributors 
responsible for the PA wave. $g_{-4}$ corresponds to the frequency band 
at 1.5 MHz. On the other hand $g_{-6}$ does to the band at 0.4 MHz; the 
frequency is distinctive from the PZT resonant frequency of 1.97 MHz, 
that is, the ringing just stems from multiple reflection of acoustic wave 
in the PZT. Thus, by removing the electric noise and 
ringing components, the PA component in time domain 
is found to be superposition of a few frequency bands. 

At the end of PA-wave analysis, it is to be noted that PA wave 
is almost invariant in 150--300 K. Therefore PA wave at 296 K was 
analysed here. 

PA-signal amplitude (solid circle) and phase (dot) are plotted against 
temperature (150--250 K) in Fig.\ \ref{fig4}. They were measured, 
by using 10-Hz and 3.49-eV laser pulses and a PZT of resonant frequency 
of 1.97 MHz, with the raising and lowering rates of 1.2 K/min. 
The incident light was parallel to $c$ axis. 
The amplitude and phase measured with decreasing temperature indicate a 
prominent peak and dip at 198 K respectively while 
those with increasing temperature have a similar peak and dip at 231 K. 
Another prominent change in the phase, that is, the leap by two degrees 
is observed at 177 K in the data on cooling. Arrows indicate the lower 
edge of phase change. The origins of changes are discussed 
in Sec.\ \ref{discussion}. 

Tens of jumped data points 
appear for both amplitude and phase only in the data on cooling. 
The irregular data presumably originate from unexpected cracking of 
the crystal on cooling because the irregularity is far less on heating. 
Besides it appears very differently on second cooling or 
in another sample. However the prominent dips and peaks are reproduced; 
it is therefore relevant that they indicate intrinsic changes in the crystal. 

Concerning with the cracking, there are a few remarks: 
(i) The PA measurement was begun at RT and kept to 150 K. 
Before measuring PA signals on heating, the sample was cooled 
to about 120 K. It was due to the thermal controller. 
(ii) The excess cooling probably induces further cracking. In fact, the 
amplitude of PA signals on heating is about half on cooling 
(see Fig.\ \ref{fig4}). 
(iii) In the present setup of Fig.\ \ref{fig1}, PA wave travels through 
the crystal. The cracking consequently decreases measured PA signals.

\section{DISCUSSION ON PHASE TRANSITIONS\label{discussion}}
\begin{figure}
\begin{center}
\includegraphics[scale=.6]{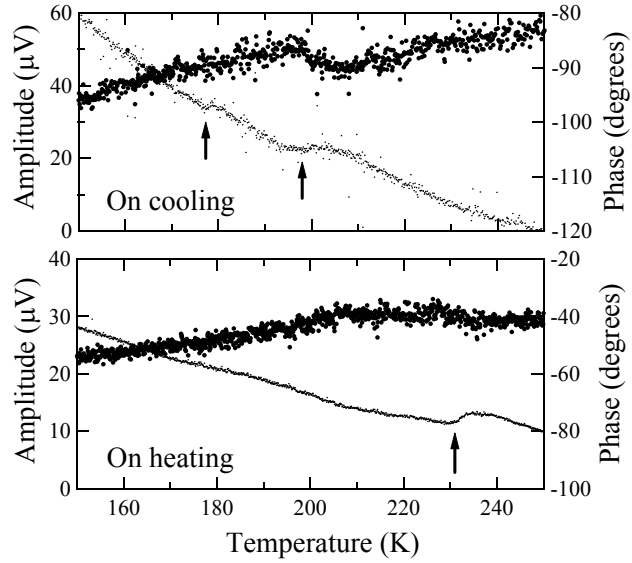}
\end{center}
\caption{\baselineskip=18pt
Amplitude (closed circle) and phase (dot) of PA signal vs 
temperature. The incident light was 10-Hz and 3.49-eV pulsed laser light 
and the incident direction was parallel to $c$ axis. 
The resonant frequency of PZT was 1.97 MHz.
The increasing and decreasing rates of temperatures were 1.2 K/min. 
Arrows indicate the lower edge of phase change and are indicative of phase 
transitions; the discussion on phase transitions is given in 
Sec.\ \ref{discussion}.}
\label{fig4}
\end{figure}

The changes of PA-signal amplitude and phase in Fig.\ \ref{fig4} 
is discussed in this section. As shown in Figs.\ \ref{fig2} and 
\ref{fig3}, PA wave consists of thermoelatic ultrasonic 
components. Therefore the amplitude and phase detected by 
a two-phase lockin amplifier correspond to those of thermoelatic 
wave. The amplitude is proportional to the peak intensity in 
two-phase lockin equipment. The phase generally depends on the amplitude. 
Figure \ref{fig4} shows the amplitude 
(closed circles) and phase (dots) of thermoelastic wave vs temperature. 
The intensity has tens of microvolts at 150--250 K 
while the phase increases as temperature becomes lower. The 
gradual changes come from the thermal deformation of crystal. 

The changes at 198 and 231 K in Fig.\ \ref{fig4} 
are probably due to those of propagation of bulk thermoelastic 
wave; the change seems different from gradual thermal deformation 
of the crystal and from the change of the PZT which has Curie 
point at 603 K. 
The changes are therefore ascribed to a phase transition at the 
temperatures. Indeed they show similar profiles 
except for irregular data. 
The transition has large hysteresis between cooling and heating. 

A theoretical study on photoacoustic 
detection of phase transition \citep{Etxe} showed that the profiles of 
amplitude and phase of PA signals are similar to those at a 
second-order phase transition. It is therefore likely that 
the transition at 198 and 231 K is second-order. 
Arrows in Fig.\ \ref{fig4} indicate the lower edge of phase 
and correspond to the transition points estimated from the indication 
by \citet{Etxe}. 
According to the calculation, the amplitude decreases 
to zero at a first-order transition because of ultrasonic 
attenuation of structural transition. 

As noted in Sec.\ \ref{intro}, phase transitions below RT in CsPbCl$_3$ 
have been hard to be detected. This is because structural study by 
x-ray diffraction is ineffective in the crystal \citep{Sakuma}. 
The result is consistent with the assignment that the type of 
transition at 198 and 231 K is second-order. 
In addition another phenomenological analysis for second-order phase 
transition \citep{Landau} derives the leap of sound velocity 
and the increase of absorption coefficient of ultrasonic wave at 
second-order phase-transition point when lattice distortion couples with 
the square of order parameter; it is often called Landau-Khalatnikov 
mechanism. The dips of amplitude in Fig.\ \ref{fig4} are explained by 
the increase in absorption coefficient of MHz-thermoelastic wave 
in the crystal. 

On the other hand another possibility for the type of transition 
still exists. The study employing x-ray diffraction indicated that 
the crystal, keeping the crystallographic group, varies the length 
of unit cell along $a$ and $c$ axes 
while it hardly changes the length along $b$ axis \citep{Sakuma}. 
It is thus possible that slight displacement of lattice 
takes place in the unit cell; the slight change might 
not be well discriminated in the x-ray-diffraction study. 
Recently order-disorder transition was reported at 193 K 
by Raman scattering \citep{Carabatos}; the assignment is consistent 
with the possibility of slight structural change.  
Consequently there are two possibilities at present; 
the type of transition at 198 and 231 K is due to 
second-order or the slight structural transition in the unit cell. 

The leap of phase at 177 K in Fig.\ \ref{fig4} appears on cooling 
and is sharper than the change at 198 K. 
It is improbable that the sudden change at 177 K comes from the adiabatic 
deformation of the crystal. That is, it also suggests the existence of 
a phase transition at the temperature. The transition is most likely 
different from that at 198 K because the profiles are inconsistent with 
each other. 
The leap at 177 K is possibly due to a second-order phase transition 
though the type is not conclusive just like that at 198 K. 

The leap at 177 K is observed only on cooling. The reason is presumably 
as follows: as remarked at the end of Sec.\ \ref{results}, 
the amplitude of PA signal on heating is two-times weaker than 
that on cooling, because of cracking. 
It is consequently harder to detect the slight change of PA signal on 
heating. Indeed the leap at 177 K on cooling is the same order with 
the noisy fluctuation of signals. Thus the leap on heating may be hidden. 

The cracking perhaps connects to the ferroelastic nature at low 
temperatures. Ferroelastic domains are typically several micrometers 
in various materials. The boundary of domains prevents 
PA wave from propagating for long distance up to 1 mm. 
On the other hand PL properties mainly depend 
on the local structure within several tens of nanometers. 
Indeed PL shows hysteresis on cooling and heating. 
The explanation for the absence of PA-signal leap on heating 
is compatible with the change of PL properties. 

In conclusion, 
PA signals in CsPbCl$_3$ crystals have been studied below RT; 
it has been shown that the leap of the PA signals 
appears at 177 and 198 K on cooling and at 231 K on heating. 
Though the profiles imply second-order phase transition, 
the slight structural transition in the unit cell is not excluded. 
Thus the types are not yet conclusive at present. 
However phase transitions below RT are confirmed by examining PA signals. 
The transition at 177 K is probably responsible for the leap of EPR signals 
\citep{Cohen} and for the sudden 
change of optical and PL properties \citep{Hayashi}.

\section*{\itshape Acknowledgments}
I would like to acknowledge Dr.\ K.\ Yoshino (Miyazaki Univ.) 
for introducing me to PAS, and also thank Mr.\ T.\ Kobayashi 
for the growth of CsPbCl$_3$ crystals and Prof.\ T.\ Hayashi (Kyoto Univ.) 
for support to the study. 
This work was partly supported by Grant-in-Aid for Research Fellow 
of the Japan Society for the Promotion of Science.

\newpage

\newpage

\newpage

\newpage

\end{document}